# Measurement of non-monotonic Casimir forces between silicon nanostructures

L. Tang[1], M. Wang[1], C. Y. Ng[1], M. Nikolic[2], C. T. Chan[1], A. W. Rodriguez[2] and H. B. Chan[1*]

[1]Department of Physics, the Hong Kong University of Science and Technology, Clear Water Bay, Kowloon, Hong Kong, China,

[2]Department of Electrical Engineering, Princeton University, Princeton, New Jersey 08544, USA;

* E-mail: hochan@ust.hk

**Casimir forces are of fundamental interest because they originate from quantum fluctuations of the electromagnetic field[1]. Apart from controlling the Casimir force via the optical properties of the materials[2-11], a number of novel geometries have been proposed to generate repulsive and/or non-monotonic Casimir forces between bodies separated by vacuum gaps[12-14]. Experimental realization of these geometries, however, is hindered by the difficulties in alignment when the bodies are brought into close proximity. Here, using an on-chip platform with integrated force sensors and actuators[15], we circumvent the alignment problem and measure the Casimir force between two surfaces with nanoscale protrusions. We demonstrate that the Casimir force depends non-monotonically on the displacement. At some displacements, the Casimir force leads to an effective stiffening of the nanomechanical spring. Our findings pave the way for exploiting the Casimir force in nanomechanical systems using structures of complex and non-conventional shapes.**

The prediction of the attraction between two neutral perfect conductors by Casimir[1] was obtained by considering the boundary conditions imposed by two planar surfaces on the quantum fluctuations of the electromagnetic field. Alternatively, the Casimir force is sometimes considered as an extension of the van der Waals force between fluctuating dipoles to solid bodies and in the regime of retardation. This attractive force increases monotonically when the distance between the two planes decreases. As the Casimir force becomes the dominant interaction between electrically neutral surfaces separated by nanoscale gaps, they are of practical importance in nanomechanical devices[16,17]. In experiments, one of the flat surfaces is often replaced by a spherical body[4-11] due to the difficulty in maintaining parallelism at small separations[18]. By introducing corrugations to one of the surfaces, recent experiments[19,20] have demonstrated the non-trivial geometry dependence of the Casimir force. Measuring the Casimir force in configurations where corrugations are present on both surfaces[21] has proven challenging due to difficulties in alignment, even when computational advances have made it possible to calculate the Casimir force between bodies of arbitrary shapes and various dielectric properties[22-26]. A number of non-conventional configurations have been predicted to yield repulsive Casimir forces[27] between objects separated by a vacuum gap. Examples include a glide-symmetric geometry[12] and an elongated metallic particle approaching a metallic plane with a circular hole[13]. For the former, the effective repulsion originates from the attractive Casimir force between components of the two interacting bodies that interpenetrate. Experimental realization of these novel designs could open the possibility for reducing stiction and levitating nanomechanical devices[6].

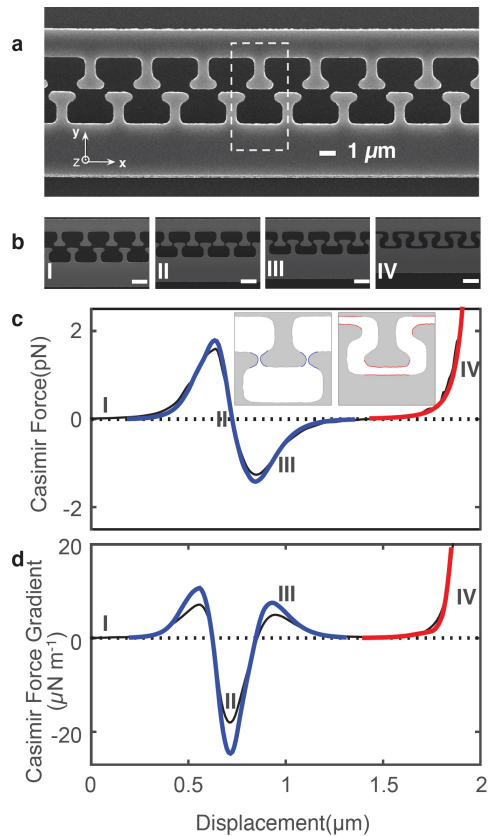

**Figure 1 | Geometry of interacting surfaces designed to generate non-monotonic Casimir forces. a,** Top view scanning electron micrograph of arrays of T-shaped protrusions at zero displacement of the movable electrode at the

lower part of the figure. The dashed line encloses one unit cell. The z axis is out of the page. **b**, Different regimes of interaction corresponding to various displacement of the bottom movable electrode along the y direction towards the top beam. The top beam vibrates with a small amplitude but the time-averaged displacement is zero. The scale bars represent 2 μm. **c**, The calculated Casimir force per unit cell along y as a function of displacement (see Supplementary Information for the procedure of weighted average). For perfectly symmetric structures, the Casimir force has no component in the x direction for all separations. The red/blue lines represent the Casimir force calculated assuming PFA for regions outlined in red/blue in the right/left inset (while neglecting all other regions). The black line represents SCUFF (boundary elements method) calculations (see Supplementary Information). Note that only numerical calculations are shown here. Measurement will be shown in Fig. 4. The insets show digitization of a typical unit cell. **d**, The gradient of the force.

Here, we design and fabricate silicon nanomechanical components with arrays of T-shaped protrusions that yield a non-monotonic Casimir force as they approach each other. The nanoscale features are defined by lithography and subsequently etched into the silicon[15]. They are hence automatically aligned after fabrication. An integrated comb actuator controls the distance between the interacting surfaces. The force gradient is detected by a vibrating silicon beam. Our experiment represents the first step towards exploiting the Casimir force between nanomechanical components with complex, non-conventional shapes.

Figure 1a shows the T-shaped protrusions with opposite orientation that are present on the two interacting bodies. The lower body approaches the top one along the y direction. This geometry is chosen because the Casimir force (also in the y direction) on the bodies is expected to exhibit strong non-monotonic dependence on the displacement. For an intuitive understanding of the origin of the non-monotonic behavior, we apply the proximity force approximation (PFA)[2] to estimate the Casimir forces between different parts of the T-protrusions and the supporting frame (Fig. 1c). At position I, the Casimir force is initially attractive as the two surfaces are far apart, with a magnitude that increases as the separation decreases. As the top of the T-shaped protrusions on the two sides approach each other, the force gradient changes sign and the Casimir force drops to zero close to the position of perfect alignment (position II). Considering only the side of the T protrusions (outlined in blue in the left inset of Fig. 1c), for small deviations from position II the Casimir force contains a y-component that acts to restore the top of the protrusions to the aligned position. The force gradient attains the most negative value at position II (Fig. 1d). Beyond position II, the Casimir force changes sign, pulling the two objects apart. At the largest displacements (position IV), the interaction is dominated by the normal Casimir force between the top of the protrusions and the main supporting frame of the other object. While the above procedure of breaking up the structure into multiple parts and applying PFA provides an intuitive understanding of the non-monotonic distance dependence, the force estimated with PFA differs quantitatively from calculations of the Casimir force using exact methods (black lines in Figs. 1c and 1d).

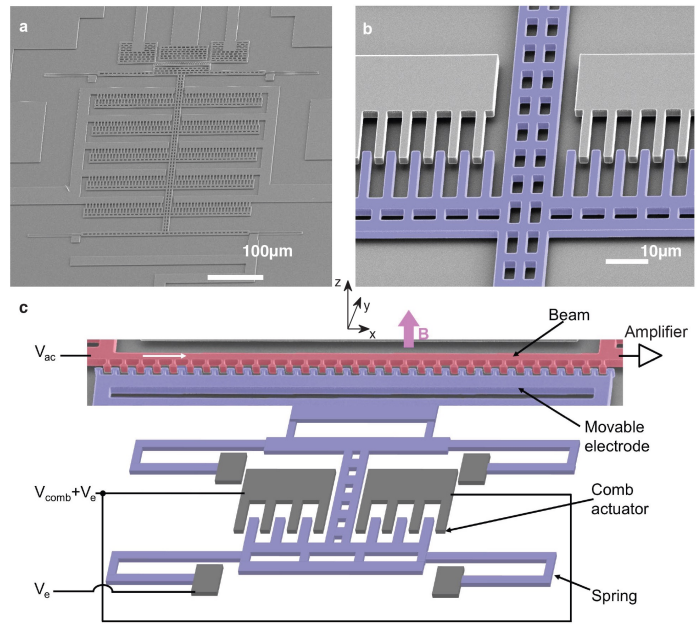

**Figure 2 | Detection and actuation scheme.** Scanning electron micrographs of **a,** the entire device and **b**, part of the comb actuator. **c**, Schematic of the detection and actuation scheme. Grey areas are fixed to the substrate via an

underlying silicon oxide layer. Blue areas are part of the movable comb. The top part is a colorized scanning electron micrograph of the beam (red) and the movable electrode (blue). There are 31 and 32 units of protrusions on the beam and the movable electrode respectively. The beam has a width of 1.5 μm (at regions away from the protrusions) and a length of 100 μm. It vibrates with a small amplitude and serves as the force gradient detector. The white arrow represents an ac current through the beam. The magnetic field B is perpendicular to the substrate.

For conventional force measurement schemes,imaging the main experimental difficulty arising from surfaces of such complex shape is the alignment when the two structures are brought into close proximity. To circumvent such difficulties, we utilize a monolithic on-chip platform[15] with integrated actuators and force sensors. As shown in Fig. 2a, the entire structure is made from the device layer of a silicon-on-insulator wafer. For the two interacting bodies in Fig. 1a, the top one is part of a doubly-clamped beam that serves as the force gradient detector (red component in Fig. 2c). The bottom structure, which we call the movable electrode (blue component in Fig. 2c), is connected to a comb actuator (Fig. 2b) that controls the distance. A potential difference $V_{comb}$ between the fixed combs (grey) and the movable combs (Fig. 2c) produces an electrostatic force along y that pushes the movable electrode towards the beam until the electrostatic force is counterbalanced by the restoring force of the supporting springs.

Figure 3a shows the measured frequency response of the beam (see Methods for the detection scheme). External forces on the beam along the y direction, including electrostatic and/or Casimir forces, lead to a change in $\omega_R$ proportional to the gradient of the total force $F$:

$$F'(d) = k \Delta\omega_R, \qquad (1)$$

where $k$ is a constant. The comb actuator produces a displacement $d$ that is proportional to the square of the potential difference $V_{comb}$ between the stationary and movable combs:

$$d = \alpha V_{comb}^2, \qquad (2)$$

where $\alpha$ is a constant. Both $k$ and $\alpha$ are calibrated by the electrostatic force generated by the applied voltage $V_e$ between the beam and the movable electrode. The electrostatic force gradient takes the following form:

$$F_e'(d) = \beta (d) (V_e - V_o)^2, \qquad (3)$$

where $\beta(d)$ is determined by the distance dependence of the electrostatic force gradient (calculated by finite element analysis) and $V_o$ is the residual voltage. Figure 3b shows the measured $\Delta\omega_R$ as a function of $V_e$ for a few different values of $V_{comb}$ (leading to different $d$). At a fixed $d$, the contributions of the electrostatic force gradient towards $\Delta\omega_R$ is parabolic in $V_e - V_o$, with a prefactor $\beta(d)/k$ that depends on the geometry. The magnitude and the sign of $\beta(d)$ determines respectively the curvature and the orientation of the parabolic dependence at fixed $d$. Figure 3c shows a surface plot of $\Delta\omega_R$ as a function of both $d$ and $V_e$. At each $d$, the voltage corresponding to the extremum of the parabolic dependence gives $V_o$. $V_o$ is found to show a weak dependence on $d$, ranging from -16 mV to -58 mV over the full range of displacement (Supplementary Information). The electrostatic contributions to $\Delta\omega_R$ are used for calibrating the measured frequency shift to the force gradient, while the vertical offset is compared to the Casimir force gradient.

From finite element simulations using COMSOL, the electrostatic force gradient on each protrusion unit attains a minimum when the top of the protrusions on the beam and the movable electrode are aligned. Taking into account contributions from different units (Supplementary Information), the displacement $d_1$ at which the minimum electrostatic force gradient occur is determined to be 772±14 nm and the proportionality constant $\alpha$ is calculated to be $5.48 \pm 0.11$ nm V$^{-2}$. The other constant $k$ is determined to be $1.07\times10^{-6} \pm 1.4\times10^{-8}$ N s rad$^{-1}$ m$^{-1}$ by a least square fit of the measured $\Delta\omega_R$ as a function of $d$ to the calculated electrostatic force gradient, as shown in Fig. 3d. Even though the finite element calculation is expected to be accurate, there exists significant uncertainty in the shape of the structures used in the calculations. Uncertainty in the digitization of the top view scanning electron micrograph and the imperfect sidewalls likely give rise to the discrepancy between measurement and calculations in Fig. 3d.

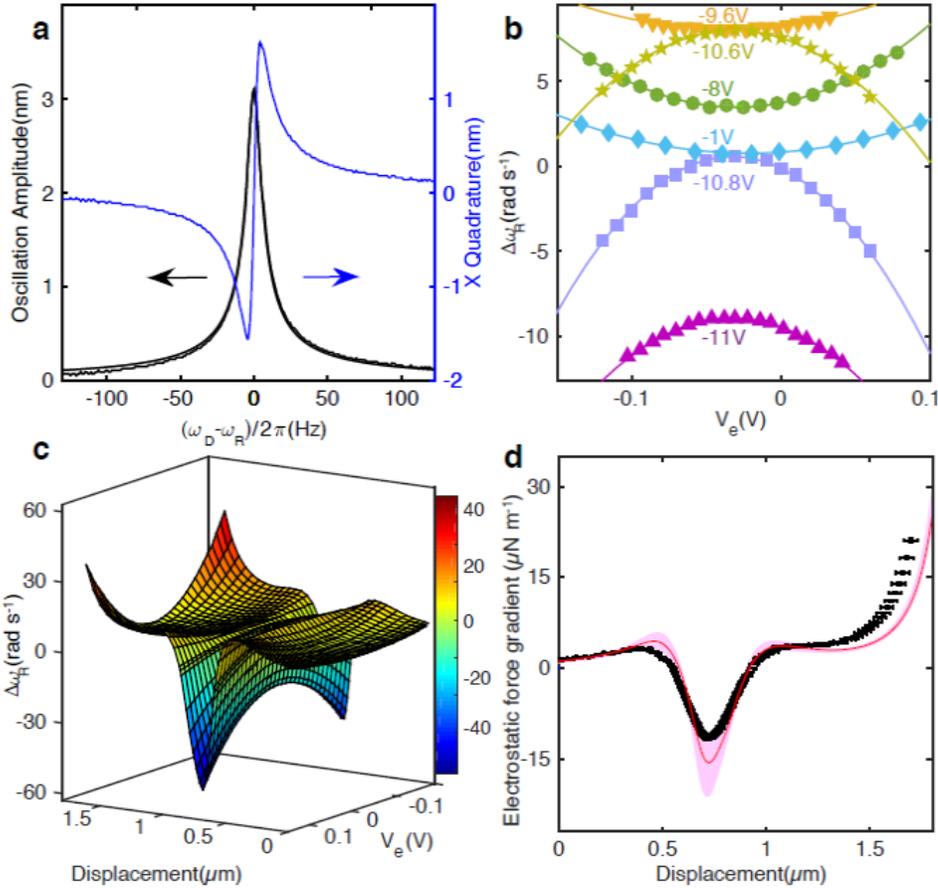

**Figure 3 | Calibration by electrostatic force. a,** Mechanical resonance of the fundamental mode of the beam, with $\omega_R/2\pi$ = 1,212,849.5 Hz and quality factor $Q$ = 58600, measured at 4 K and < $10^{-5}$ Torr. The procedure to infer the shifts in $\omega_R$ from the measured X quadrature of vibration is explained in Methods. **b,** Measured shift in the resonant frequency of the beam as a function of applied voltage $V_e$ between the beam and the movable electrode for different values of $V_{comb}$. The values of $V_{comb}$ are labelled next to the curves. Solid lines are parabolic fits. **c,** Measured $\Delta\omega_R$ as a function of displacement and $V_e$. **d,** Measured electrostatic force gradient at $V_e = V_0 + 100$ mV, where $V_o$ represents

the residual voltages at which the extrema of the parabolas (similar to those in **b**) occur. The error bars for the force gradient are calculated from electrical noise in recording the X quadrature of vibrations of the beam. The error bars in displacement originates from the uncertainty in locating the displacement at which the electrostatic force gradient attains minimum. The red line is a least square fit using electrostatic force calculations from finite element analysis. For a rough estimate of the uncertainty in the shape of the structure used in the calculations, the pink band shows the range of electrostatic force obtained when the digitized geometry is expanded or shrunk by one pixel of the top view scanning electron micrograph.

We minimize the electrostatic contribution by setting $V_e = V_o\,(d)$ and measure the force gradient between the two structures as $d$ increases. Figure 4 shows the measured results, plotted in black. The solid red line shows the calculation of the Casimir force for the same geometry as measured in experiment, with no fitting parameters. The calculations are performed with SCUFF-EM, an open-source software implementation of the boundary-element method[28] that meshes the surface of the interacting bodies and calculates the Casimir energy using the so-called fluctuating-surface-currents method[22]. In the calculations, the finite conductivity of the silicon is taken into account and the temperature is assumed to be zero. We checked that at the temperature of measurement (4 K), the thermal correction is negligible. (At $d = 1.83$ μm, the zeroth Matsubara frequency term contributes to < 0.3% of the force.) Overall, the calculation reproduces all of the main features of the experiment, including the two maxima, the minimum and the sharp increase of the force gradient. The measured and calculated values are mostly in good agreement. A number of factors likely contribute to the discrepancy, such as the variation of the geometry between different unit cells (supplementary information), contribution of patch potentials, imperfect sidewalls and uncertainties in digitizing the micrograph. For instance, expanding the size of the structures by one pixel of the micrograph (~ 5 nm) normal to the digitized boundary leads to nearly a factor of 2 increase in the calculated force on a unit cell.

Given the complex shape of the structures and the difficulty in characterizing the geometry, we consider our measurement to be in good agreement with the calculations of the Casimir force.

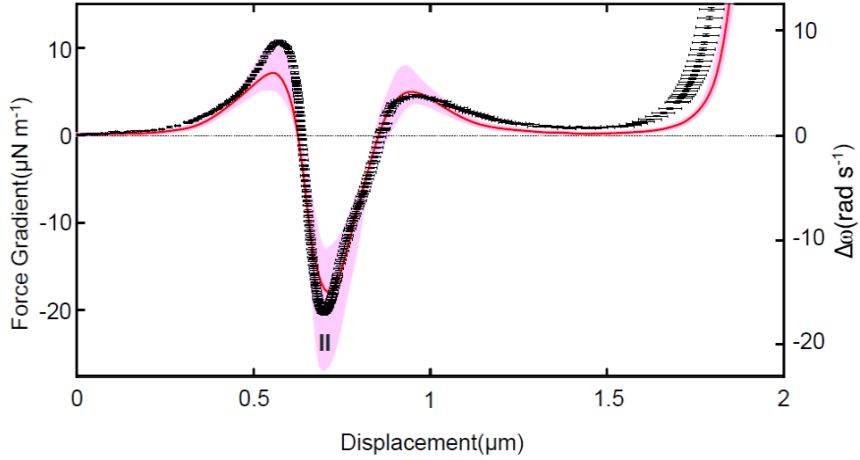

**Figure 4 | Measured force gradient per unit cell after compensating for residual voltage.** Measured data are plotted as dots with error bars. The red line represents the Casimir force calculated with boundary elements method, with the shape of the structures digitized from the top view SEM. Non-uniformities in the shapes of different T-protrusions are taken into account (Supplementary Information). Finite conductivity of the silicon is also included (see Methods). The pink band around the red line represents the change in the calculated Casimir force when the digitized geometry is expanded or shrunk by one pixel of the top view scanning electron micrograph. The error bars for the force gradient are calculated from electrical noise in recording the X quadrature of vibrations of the beam. The error bars in displacement originates from the uncertainty in determining the displacement at which the electrostatic force gradient in Fig. 3d attains minimum. The "Casimir spring" effect attains maximum at position II (following the notation in Fig. 1d).

There are a number of interesting features in Fig. 4. First, the force gradient changes sign at two locations ($d \sim 0.6$ μm and 0.8 μm), indicating that the Casimir force depends non-monotonically on displacement. To our knowledge, non-monotonic Casimir forces have not been measured in previous experiments. Second, when the top of the protrusions are aligned, the resonant frequency (shown as the right vertical axis of Fig. 4) of the beam exceeds the unperturbed value in the absence of Casimir and electrostatic forces. The quantum fluctuations therefore produce a "Casimir spring" effect that leads to additional confinement of the mechanical resonator, in close analogy to optical spring effects[29] due to real photons in optomechanical systems. The measured "Casimir spring" effect is, however, much weaker than our nanomechanical silicon spring or typical optical springs. Third, we compare our results to theoretical findings by Rahi et al.[30] on the stability of equilibria produced by Casimir forces. Assuming that the optical properties of the materials can be characterized by their dielectric function (with negligible magnetic susceptibility), it was shown that Casimir forces between bodies separated by vacuum can only generate equilibria that are unstable. At the minimum of the force gradient in Fig. 4, the Casimir force tends to restore the structures back to the position in which the top of the protrusions are aligned. However, our results do not violate the aforementioned theorem, because the system is unstable with respect to small displacements in the x direction perpendicular to the direction in which the force gradient is detected. Restoring forces in the x direction are provided by the springs supporting the movable electrode and the combs.

In summary, we have shown that the Casimir force between two silicon structures with T-shaped protrusions depends non-monotonically on the displacement. The Casimir force confines the motion along the measurement direction, leading to effective stiffening of the mechanical spring. Our findings pave the way for exploiting the Casimir force in nano-machinery consisting of components of unconventional geometries. The experimental scheme can also be used to study lateral Casimir forces between structures of other novel shapes.

**Methods**

**Device fabrication.** The device is created from a silicon-on-insulator wafer (Supplementary Information) with the p-doped silicon device layer and the buried oxide layer having thickness of 2.23 μm and 2.0 μm respectively. The van der Pauw method is used to measure the resistivity and carrier concentration of the device layer to be $4.23 \times 10^{-3}$ Ω cm and $2.2 \times 10^{19}$ cm$^{-3}$ respectively at 4K.

Ultraviolet stepper lithography is used to create all the components, including the beam, movable electrode and comb actuator. The resist pattern is transferred to a silicon oxide etch mask that is grown on the device layer. With the etch mask protecting the silicon underneath, the unwanted silicon is removed by deep reactive ion etching. Hydrofluoric acid is then used to remove the oxide etch mask and undercut the buried oxide to give suspended mechanical components. The surfaces that interact via the Casimir force are therefore automatically aligned after fabrication.

The hydrofluoric acid passivates the surfaces of the silicon components and prevents oxidation of the silicon at ambient pressure for a few hours. After completion of the fabrication process, the device is immediately placed in a sealed probe and evacuated to pressure of ~10$^{-5}$ Torr. The probe is then inserted into a cryostat at 4K with a magnetic field of 5T.

**Measurement of $\Delta\omega_R$.** The arrow in Fig. 2c illustrates an ac current that, in the presence of a magnetic field (in the z direction), generates a periodic Lorentz force on the beam in the y direction. The frequency of the ac current is chosen to excite the fundamental mode of the beam, which vibrates parallel to the substrate. Motion of the beam in the magnetic field produces an electromotive force that reduces the amplitude of the current by an amount proportional to the vibration amplitude. $\Delta\omega_R$ is measured by setting $V_{ac}$ to a pre-determined frequency $\omega_{ref}$ and recording the X quadrature of the output of the lockin amplifier. Assuming that $\omega_{ref}$ is close to $\omega_R$, the deviation of $\omega_R$ from $\omega_{ref}$ can be determined by using the linear dependence of the X quadrature on the excitation frequency shown in Fig. 3a. Vibration amplitude of the beam is about 3 nm. The calibration involves increasing the periodic current until the beam vibrations reach the nonlinear regime. Just before the onset of bistability, the critical vibration amplitude is determined by the beam dimensions that are measured in a scanning electron microscope.

**Calculating the electrostatic force and the Casimir force.** Calculations of the forces are based on digitization of the top view scanning electron micrographs. The structures are assumed to be invariant in the z direction (normal to the substrate), rendering the geometry two dimensional. This assumption is justified because for all the major features of the Casimir force considered here, the interacting elements of the T-protrusions are within about 200 nm of each other (panels II and IV of Fig. 1b), while the thickness of the structure is more than 10 times larger.

For the dielectric function $\varepsilon(i\xi)$ that is required in the Casimir force calculations, , the following expression is used[31]:

$$\varepsilon(i\xi) = 1.035 + (11.87 - 1.035)/(1 + \xi^2/\omega_0^2) + \omega_p^2/[\xi(\xi + \Gamma)], \qquad (4)$$

where $\omega_0 = 6.6\times10^{15}$ rad s$^{-1}$, $\omega_p = 4.53\times10^{14}$ rad s$^{-1}$ and $\Gamma = 7.69\times10^{13}$ rad s$^{-1}$. The last term in Eq. (4) accounts for contributions from extra carriers introduced via doping[32]. $\omega_p$ and $\Gamma$ are deduced from the measured resistivity of silicon ($4.2\times10^{-3}$ Ω cm)[33], with $m^* = 0.34\ m_e$ for the effective mass of electrons in p-doped silicon.

The PFA was applied in two different ways to estimate the Casimir force, as illustrated in the two insets in Fig. 1c. For the red lines in Fig. 1c and Fig. 1d, calculations follow the conventional approach, by dividing the interacting surfaces (the red parts in the right inset of Fig. 1c) into pairs of parallel plates of small area, facing each other in the y direction and then summing up the force between the pairs of plates. The good agreement between calculations using SCUFF-EM and PFA (the red and black lines in Fig. 1c and Fig. 1d) is somewhat expected, as the interacting components resembles parallel plates. For the blue lines in Fig. 1c and Fig. 1d, we follow the procedure for calculating lateral Casimir forces in previous experiments[21], where the interacting surfaces (the blue parts in the left inset of Fig. 1c) are divided into parallel plates that face each other in the x direction. With a known separation between each pair of parallel plates, the Casimir energy (instead of the force in the former case) between them is calculated ignoring any fringe effects. The total Casimir energy $E$ at a fixed displacement $d$ is then obtained by summing over all pairs of plates. After repeating this procedure for the entire range of $d$ to obtain $E(d)$, the Casimir force is calculated by taking spatial derivative (along the y direction) of $E(d)$ with respect to $d$. As shown by the blue line in Fig. 1d, the PFA over-estimates the Casimir force gradient by about 30%. Similar deviations of the PFA from the Casimir force have been previously observed in lateral Casimir force measurements in sinusoidal gratings[21].

For the measured Casimir force gradient in Fig. 4, even though the values generally agree well with SCUFF, deviations reach ~30% at 0.55 μm. Therefore our measurement does not provide unambiguous evidence of the breakdown of PFA.

**Acknowledgements**

H.B.C., L.T. and M.W. are supported by HKUST 16300414 from the Research Grants Council of Hong Kong SAR. C.Y.N and C.T.C are supported by AoE/P-02/12 from the Research Grants Council of Hong Kong SAR. M.N. and A.W.R. are supported by National Science Foundation under Grant no. DMR-1454836.

**Author contributions**

L.T. and M.W. fabricated the devices and conducted the measurements. C.Y.N., M.N., A.W.R. and C.T.C. performed the theoretical calculations. H.B.C. conceived and supervised the experiment. All authors discussed the result and contributed to the writing.

# Measurement of non-monotonic Casimir forces between silicon nanostructures

# Supplementary information

L. Tang[1], M. Wang[1], C. Y. Ng[1], M. Nikolic[2], C. T. Chan[1], A. W. Rodriguez[2] and H. B. Chan[1*]

[1]Department of Physics, the Hong Kong University of Science and Technology, Clear Water Bay, Kowloon, Hong Kong, China,

[2]Department of Electrical Engineering, Princeton University, Princeton, New Jersey 08544, USA;

**Device fabrication**

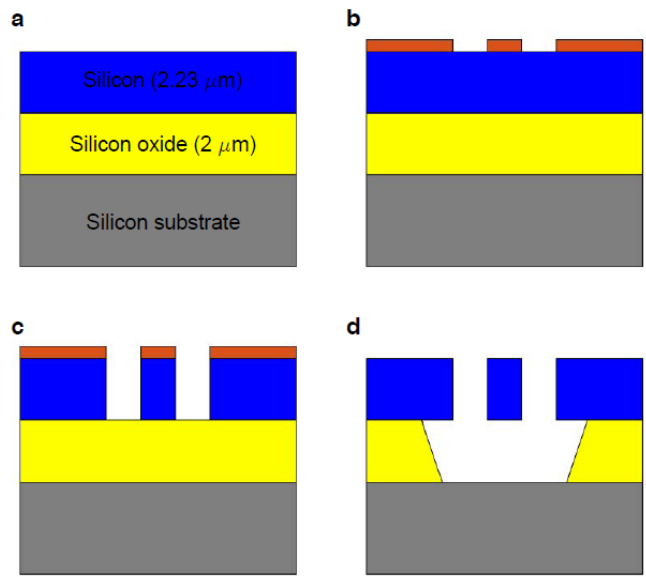

**Supplementary Figure S1 | The fabrication procedure of the device (not to scale). a**, A cross-sectional view of the silicon-on-insulator wafer. The silicon device layer, the buried oxide layer and the substrate are shown in blue, yellow and grey respectively. **b**, The silicon oxide etch mask (red) is created using the resist pattern from lithography. **c**, Silicon in the regions not protected by silicon oxide is removed by DRIE. **d**, HF selectively etches the silicon oxide isotropically, undercutting of the top silicon structure by ~ 2.7 μm. The middle silicon piece is thin enough to be suspended. The other two pieces have oxide underneath and therefore are anchored to the substrate.

The device is created from a silicon-on-insulator wafer with the p-doped silicon device layer and the buried oxide layer having thickness of 2.23 μm and 2.0 μm respectively (Fig. S1). A layer of thermal oxide (435 nm) is grown on the wafer to act as the etch mask for the underlying silicon. Ultraviolet stepper lithography is performed to define the shape of all the components, including the T-protrusions, the beam, the movable electrode and the comb actuator. The resist pattern is transferred to the oxide layer with plasma etching. With the silicon oxide mask protecting the silicon underneath, the unwanted silicon is removed by deep reactive ion etching (DRIE). To produce near-vertical sidewalls with no undulations, a continuous etch and passivation recipe is used. Afterwards, an oxygen plasma etch removes the hydrocarbon generated by the DRIE.

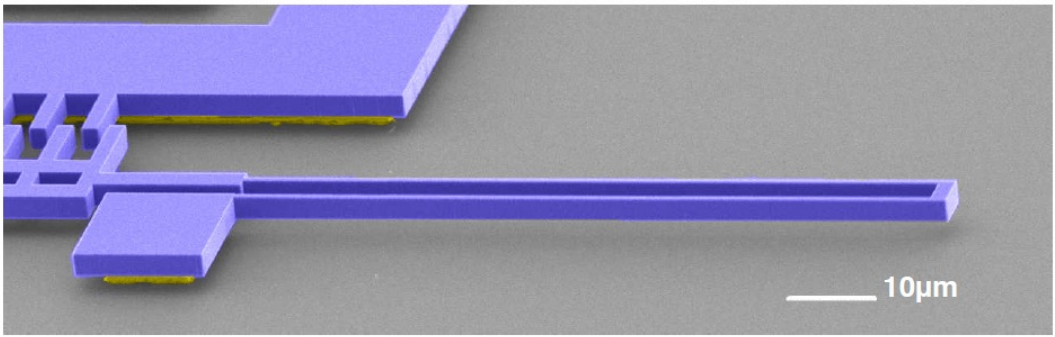

**Supplementary Figure S2 | Colorized scanning electron micrograph of a spring and its anchor**. **a**, The blue layer represents the device layer used for the silicon components. For the anchors of the spring and the fixed comb, the silicon oxide underneath (yellow) attaches the silicon structure to the substrate.

Hydrofluoric acid is then used to remove the oxide etch mask and undercut the buried oxide to give suspended mechanical components. This wet etch is isotropic and its duration is chosen so that it undercuts the silicon in the device layer by about 2.7 μm. As a result, silicon components in the device layer becomes suspended if their width is less than ~ 5 um. Figure S2 shows that the movable parts of the serpentine spring is suspended while the large anchor area is attached to the substrate by the underlying silicon oxide that remains after the wet etch. The fixed comb electrodes are also attached to the substrate in a similar fashion.

**Dependence of displacement on comb drive voltage**

While we do not have the capability to measure the displacement of the comb actuator at 4K under the exact same conditions of our force measurement, we check the quadratic dependence of the displacement of the comb actuator on $V_{comb}$ at room temperature under an optical microscope. For each value of $V_{comb}$ in Fig. S3, we record a top view image and extract the displacement. The measured displacement is well-fitted with a parabola with a minimum at 29 mV. The non-zero value of this offset voltage leads to errors in Eq. 2 that is much smaller than the horizontal error bars in Fig. 3d and Fig. 4.

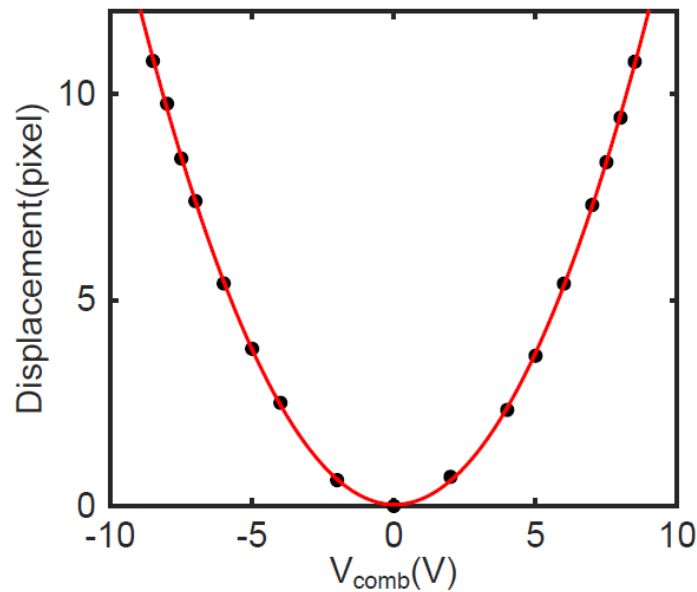

**Supplementary Figure S3 | Measured displacement of the comb actuator as a function of $V_{comb}$. a**, Each black circle represents the displacement measured from a digitized optical image. The red line is a parabolic fit.

**Comparing electrostatic force calculations from COMSOL and SCUFF-EM**

Electrostatic force calculations can also be performed with SCUFF-EM using the same mesh as the Casimir force calculations. Figure S4 compares the electrostatic force on unit 16 calculated with COMSOL and SCUFF-EM. There is good agreement between the two calculations (~1% difference at $d = 0.6$ μm).

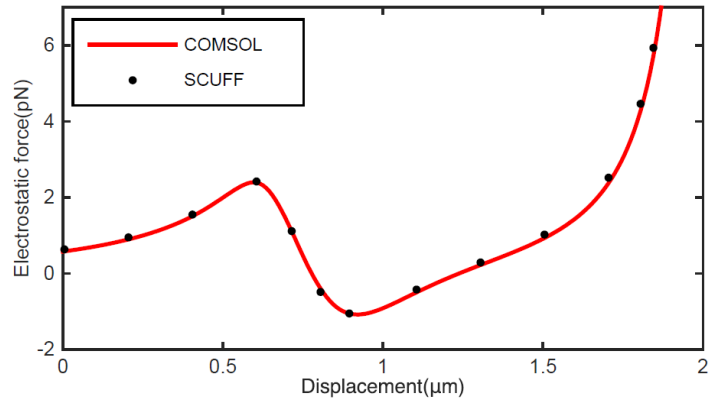

**Supplementary Figure S4 | Comparison of electrostatic force calculated with COMSOL and SCUFF-EM.** The potential difference is chosen to be 0.1 V. The electrostatic force calculated with COMSOL and SCUFF-EM on unit 16 is shown as a red line and black circles respectively.

The electrostatic calculations are effectively two dimensional, assuming that the structures extend to infinity in the z direction and ignoring the substrate. For comparison with measurements, the force per unit thickness is multiplied by the thickness of the structure to yield the calculated force on each unit. To justify using this approximation, we perform a 3D calculation with COMSOL on unit cell 16, with the substrate included. Figure S5 shows that there is good agreement between the 3D and 2D calculations.

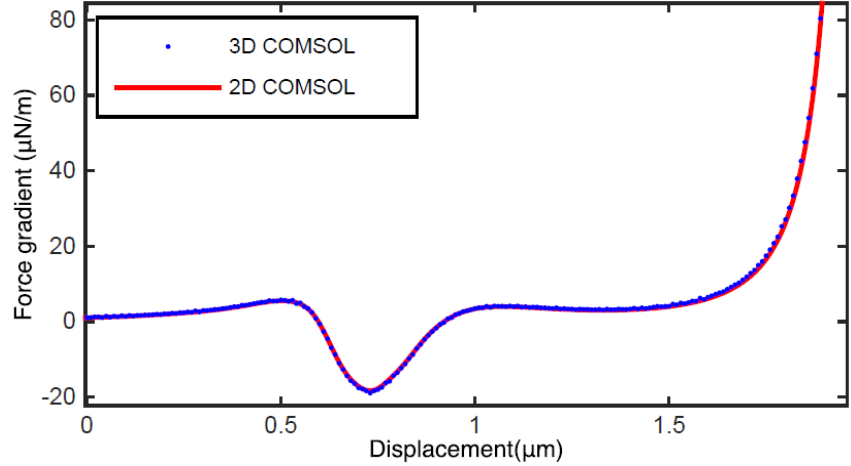

**Supplementary Figure S5 | Comparison of 2D and 3D calculations of the electrostatic force gradient calculated with COMSOL.** The potential difference is chosen to be 0.1 V, for unit 16. The 3D calculation, with the substrate included, is shown as blue points. The force per unit thickness obtained from the 2D calculations is multiplied by the thickness of the structure and differentiated to give the red line.

**Dependence of residual voltage on displacement**

Figure S6 shows the measured residual voltage $V_0$ as a function of the displacement. With the complicated shape of the silicon T-protrusions, different parts of the interacting surfaces have different crystal orientations and different work functions. It is somewhat expected that $V_0$ does not remain constant with distance.

Since the electrostatic force gradient becomes zero at around $d = 0.55$ μm and 0.9 μm, the dependence of the measured force gradient on $V_e$ becomes very weak. As a result, there are large uncertainties in $V_0$ in this range of $d$.

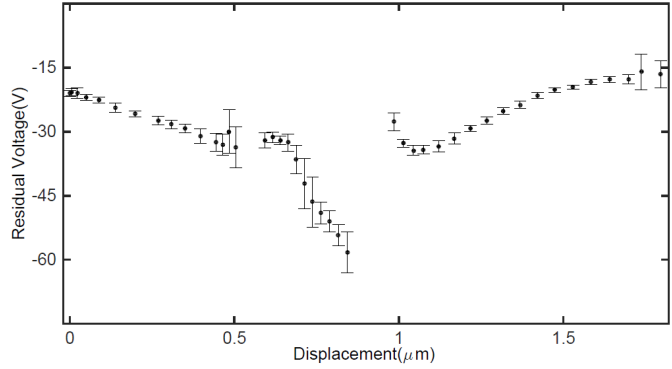

**Supplementary Figure S6 | Dependence of residual voltage on displacement.**

**Summing up the force from the units cells**

Simulating the entire structure of 31 units, each with submicron complex features, requires computational resources beyond our capability. Instead, numerical calculations of the Casimir force and electrostatic force were performed at the level of one unit cell. While correlations and effects due to multiple scattering between unit cells are bound to become important at large separations (> 2 μm, corresponding to d < -1.9 μm, where the negative sign corresponds to pulling the interacting structures further apart), our paper focuses on the interpenetrating region where such effects are negligible. The force we considered arises from components within a unit cell coming into proximity (~ 100 nm) of each other. In this regime, effects from neighboring units are negligible. To verify the validity of our approach, we perform an additional calculation of the Casimir force using two unit cells as the interacting structure. At displacement of 0.85 μm, the calculated Casimir force equals twice that on a single unit cell to within 1.6%, justifying the use of pairwise addition at the unit cell level.

**Non-uniformity in the T-protrusions**

After the lithography and etching steps, the dimensions of the T-protrusions often differ considerably from the nominal design values. It is therefore necessary to fabricate and test many samples with different dimensions. The device in this paper is chosen (out of 5 devices tested in the same fabrication run) because the Casimir force is well above our detection sensitivity and at the same time the effects of non-uniformity (described below) on the force is not excessive.

Due to non-uniformities in the lithography and etching processes, the shapes of the T-protrusions in the array are not identical. We measured the shape of every unit in a scanning electron microscope. Figure S7c shows that the measured separation between the top of neighboring protrusions on the beam and the electrode ranges from 46 nm to 85 nm. Based on the boundary elements method calculations, the Casimir forces on different units can vary by up to a factor of 5. It is therefore necessary to take such non-uniformity into account in comparing measurement to theory.

Units at different locations contribute differently to the overall Casimir force, depending on the distance of the unit from the two ends of the beam that are fixed in position. For the sake of explanation, let us assume that each unit generates the same force locally. Unit 18 near the middle of the beam excites stronger vibrations of the beam compared to unit 1 near the end, leading to a larger change in the resonant frequency. Therefore, in calculating the total Casimir force or the electrostatic force, the contributions from the 31 units need to be weighted. The weight is obtained from finite element analysis. As shown in Fig. S7, a point force at frequency $\omega_R$ is exerted at different positions along the doubly-clamped beam and the resultant vibration amplitude at the middle of the beam is calculated. As expected, the vibration amplitude attains a maximum (minimum) when the force is exerted at the center (edge). Contributions from different units to the Casimir and electrostatic force gradients are weighted by a factor that is proportional to the vibration amplitude calculated above.

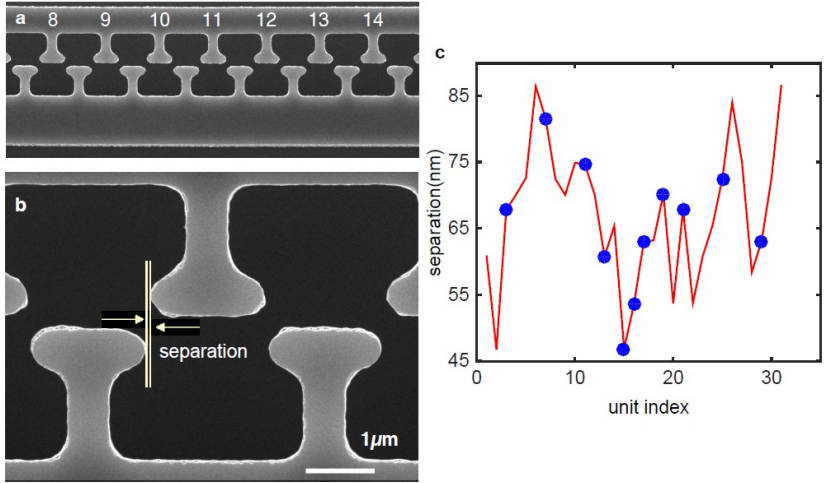

**Figure S7| Distribution of separation between T-protrusions. a.** Scanning electron micrograph of the T protrusions, units 8 to 14. **b.** Close-up to define the separation between the protrusions. **c.** Measured values of the separation for all 31 units.

Due to limited computation resources, it is not feasible for us to calculate the Casimir force for all the units. Instead, we only calculated the force on one third out of the 31 units. The units used for Casimir force calculations are marked as blue in Fig. S7c.

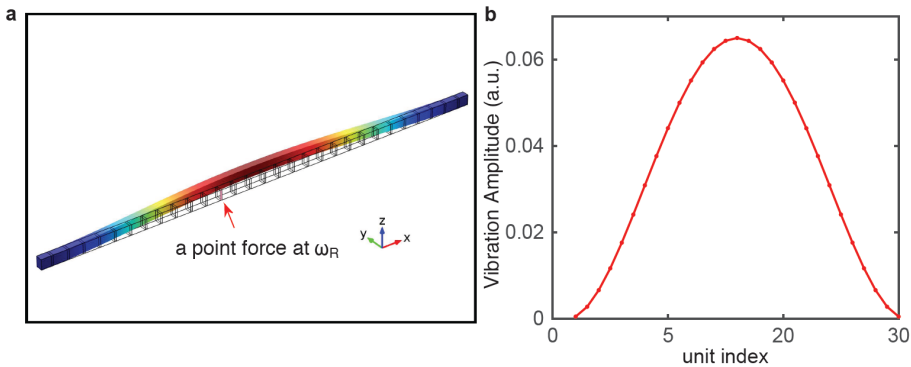

**Figure S8| Contribution of different T-protrusions on the measured force gradient. a.** COMSOL simulations of a point force with frequency $\omega_R$ applied at unit cell #13, exciting the fundamental mode of vibration. **b.** Calculated vibration amplitude of the beam when the point force is applied at different locations across the beam.

**Finite conductivity effects**

In Figure S9, the Casimir force computed by SCUFF-EM for a particular unit cell is plotted as black circles. When the silicon is replaced with perfect conductors, the Casimir force is shown as red squares. With the material properties described by Eq. (4), the Casimir force is reduced by up to ~70 %.

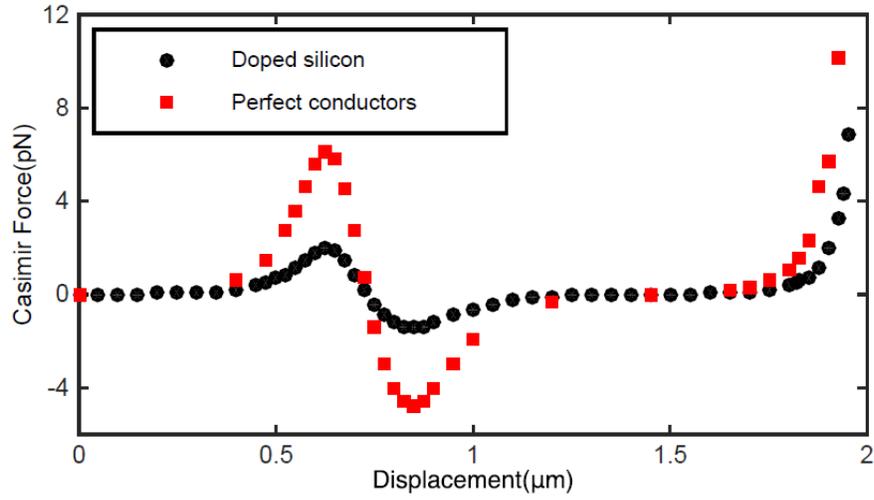

**Figure S9| Finite conductivity effects on the Casimir force.** Casimir force computed by SCUFF-EM for unit cell 16, made of silicon (black circles) and perfect conductors (red squares).